\documentclass[%
 reprint,
 amsmath,amssymb,
 aps,
pra,
]{revtex4-2}

\usepackage{graphicx}
\usepackage{dcolumn}
\usepackage{bm}
\usepackage{hyperref}

\hypersetup{
    colorlinks=true,
    linkcolor=blue,
    filecolor=magenta,      
    urlcolor=cyan,
}

\usepackage{siunitx}
\usepackage{braket}

\usepackage{placeins}

\begin{document}

\preprint{APS/123-QED}

\title{Revealing Collective Emission in the Single-to-Bulk Transition of Quantum Emitters in Nanodiamond Agglomerates}

\author{Jonas Gutsche}
\email{gutsche@rhrk.uni-kl.de}
\author{Ashkan Zand}%
\author{Marek Bültel}%
\author{Artur Widera}%
\affiliation{%
 Technische Universität Kaiserslautern und Landesforschungszentrum OPTIMAS, 67663 Kaiserslautern, Germany\\
}%

\date{\today}

\begin{abstract}
Individual quantum emitters form a fundamental building block for emerging quantum technologies. Collective effects of such emitters might improve the performance of applications even further. When scaling materials to larger sizes, however, collective effects might be covered by transitions to bulk properties. Here, we probe the optical properties of Nitrogen Vacancy (NV) centers in agglomerates of nanodiamonds. We quantify the transition from individual emitters to bulk emission by fluorescence lifetime measurements, and find a transition to occur on a length scale of $\sim 3$ wavelengths around the emitter. While our lifetime measurements are consistent with superradiant decay, the second-order correlation function, which is a standard measure to reveal collective properties, fails to probe collective effects for our case of an ensemble of collectively contributing domains to the emission. Therefore, we propose and apply a new measure to trace collective effects based on the fluctuation statistics of the emitted light. Our work points toward systematically studying collective effects in a scalable solid-state quantum system, and using them for quantum optical applications in agglomerates of highly-doped nanodiamonds. 
\end{abstract}

\maketitle

\section{\label{sec:level1}Introduction}

Micro- and nano-scale objects have become increasingly relevant for technological applications \cite{Bayda.2019, Roco.2003}. 
The length scale of a material platform is often essential to understand its properties. When its size increases, a transition occurs at a characteristic length scale to establish macroscopic bulk properties.
Prominent examples are the structure of nano-gold complexes \cite{Jadzinsky.2007}, 
aqueous solutions of gold and silver colloids \cite{Brust.2002}, 
 or the emergence of nanoscopic aqueous droplets of acid formed within a superfluid helium cluster \cite{Gutberlet.2009}. 
Impurities immersed in such a material will experience the bulk properties also beyond a particular length scale, which for optically active impurities can be of the order of the wavelength of light. 
For multiple impurities, an additional transition occurs, beyond which the impurities act collectively and develop properties that are not present in individual nano-particles, making collective effects attractive for possible applications. 
Such collective effects can  lead to various physical phenomena, such as the well-known superradiance predicted by Dicke \cite{Dicke.1954}. An ensemble of emitters emit energy collectively, yielding insight into quantum mechanical processes through a macroscopic effect \cite{Scully.2009, Gross.1982}. 
Collective emission has been observed in 
gases \cite{Skribanowitz.1973}, 
quantum dots \cite{Scheibner.2007}, 
and ultracold atoms \cite{Baumann.2010}, 
for example. 
Recently, superradiance has been observed in optically active nitrogen-vacancy (NV) centers in diamond \cite{Bradac.2017, Angerer.2018}. 
An important question for future applications is if the signature of collective effects prevails when the system size crosses the transition to bulk properties.

In this respect, the negatively charged NV center in diamond is a promising technological platform. It is a widely investigated material system \cite{Doherty.2013}, specifically for emerging quantum technology such as spin-magnetometry \cite{Maletinsky.2012}, quantum information processing \cite{Neumann.2010}, and non-classical light sources \cite{Beveratos.2001}.
This point defect is well-known as a room-temperature single-photon source \cite{Gruber.1997}
and 
can be produced as bulk- and nanodiamond \cite{Ohno.2012, Narayan.2017} and prepared 
as individual emitters or assembled as large, dense ensembles.
Likewise, these defect centers change their (quantum) optical properties as a function of the size of the diamond host crystal and the distance between individual defect centers. 
On the one hand, NV centers in nanodiamonds are subject to a reduction of their optical decay rate in comparison to bulk diamond due to the increased surface-volume ratio
and thereby reduced density of states (DOS) \cite{Beveratos.2001, Inam.2013}. 
On the other hand, nanodiamond samples with high NV-concentration and a particle size above \SI{100}{\nano\meter} show collective emission, i. e. Dicke Superradiance \cite{Dicke.1954, Bradac.2017}.
These collective effects are accompanied by a super-Poissonian photon distribution leading to an increase in the second-order correlation at zero time delay $g^{(2)}(0)>1$, the so-called photon bunching.
These properties make NV centers in nanodiamonds well-suited to study parallel transitions to bulk properties and  collective response. 

\begin{figure*}
\includegraphics[width=0.9\textwidth]{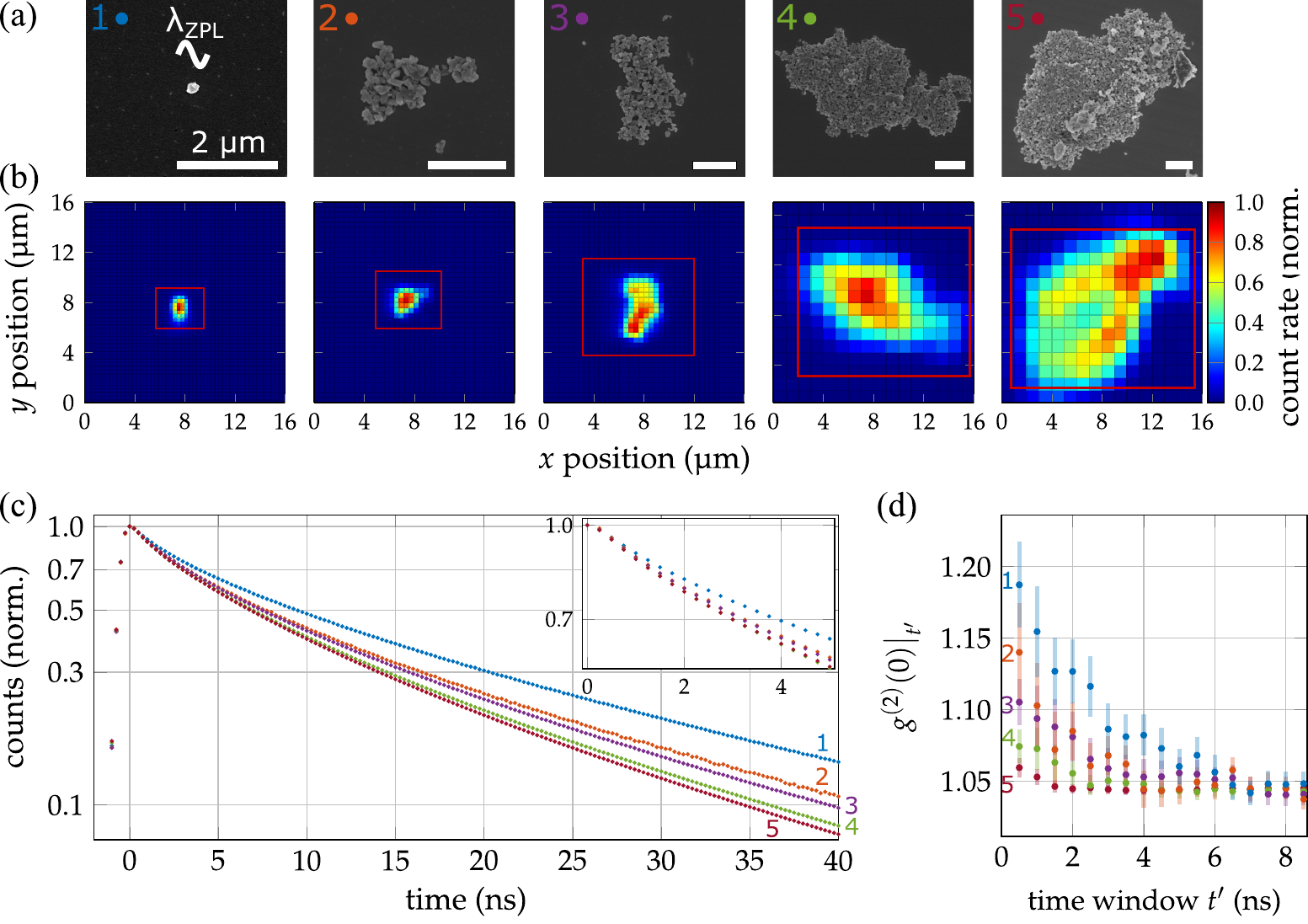}
\caption{Raster scan of nanodiamonds in different agglomeration states. \textbf{(a)} SEM images of different agglomeration states ranging from single nanodiamonds over planar sections to an agglomerate with 3D features. The scale bar in each image has a size of \SI{2}{\micro\meter}, and a scale of the emission wavelength $\lambda_{\mathrm{ZPL}}$ is shown for comparison. \textbf{(b)}  Normalized count rate measured at a 2D confocal scan as a function of $x$- and $y$-position; all scans have the same scale; the red rectangle gives the region of the respective SEM-Image. \textbf{(c)} Count rate weighted lifetime measurements of the different agglomerates shown in (a) and (b). A multi-exponential decay is observed, where larger agglomerates show a lifetime reduction, the zoom-in shows the initial fast decay. \textbf{(d)} Time-averaged second-order correlation measurements in dependence on the time window $t'$ with maximum signal to noise ratio of the shown agglomerates. The photon bunching reduces on a time scale smaller than the fluorescence lifetime.
}
\label{fig:intro}
\end{figure*}

We use agglomerates of nanodiamond crystals as an intermediate state between bulk diamond and individual nanodiamonds, where poly-crystalline quasi-2D sections form on a glass substrate. 
Each doped nanodiamond ($c_\mathrm{NV} \approx \SI{10}{ppm}$) comprises  \SI{>1000}{} NV centers, which can be concentrated at sub-wavelength scales.
Additionally, the size of agglomerates can be controlled, as shown in figure \ref{fig:intro}, where a change of their optical emission properties becomes immediately apparent.
A profound size effect on the optical properties concerning lifetime reduction in agglomerates of small nanocrystals with a size of \SI{5}{\nano\meter} was shown but not investigated in detail \cite{Smith.2009}.
 
Here, we probe the optical properties as the diamond host transits to bulk-like behavior, and we reveal the collective phenomena in this condensed-matter material. 
To this end, we investigate the dependence and scaling of (quantum) optical properties of NV-ensembles on the system's size in a systematic study of different agglomeration states ranging from single nanodiamonds of size \SI{100}{\nano\meter} to large quasi 2D agglomerates consisting of up to 10000 nanodiamonds.
We characterize the quantum optical emission properties employing fluorescence-lifetime imaging and second-order correlation measurements.
We introduce a novel way to quantify collective effects via the Fano factor, quantifying fluctuations of the photon statistics, and justify this in a theoretical model.
We furthermore identify the relevant length scale on which bulk-like properties can be established and the scaling of collective properties in agglomerates.
Our results are not restricted to NV centers but also applicable to other (diamond) color centers like silicon-vacancy (SiV) or germanium-vacancy (GeV) centers, or quantum dots. 

\section{Experimental system and measurement sequence \label{ch:setup}}

\begin{figure}
\centering
\includegraphics[width=0.4\textwidth]{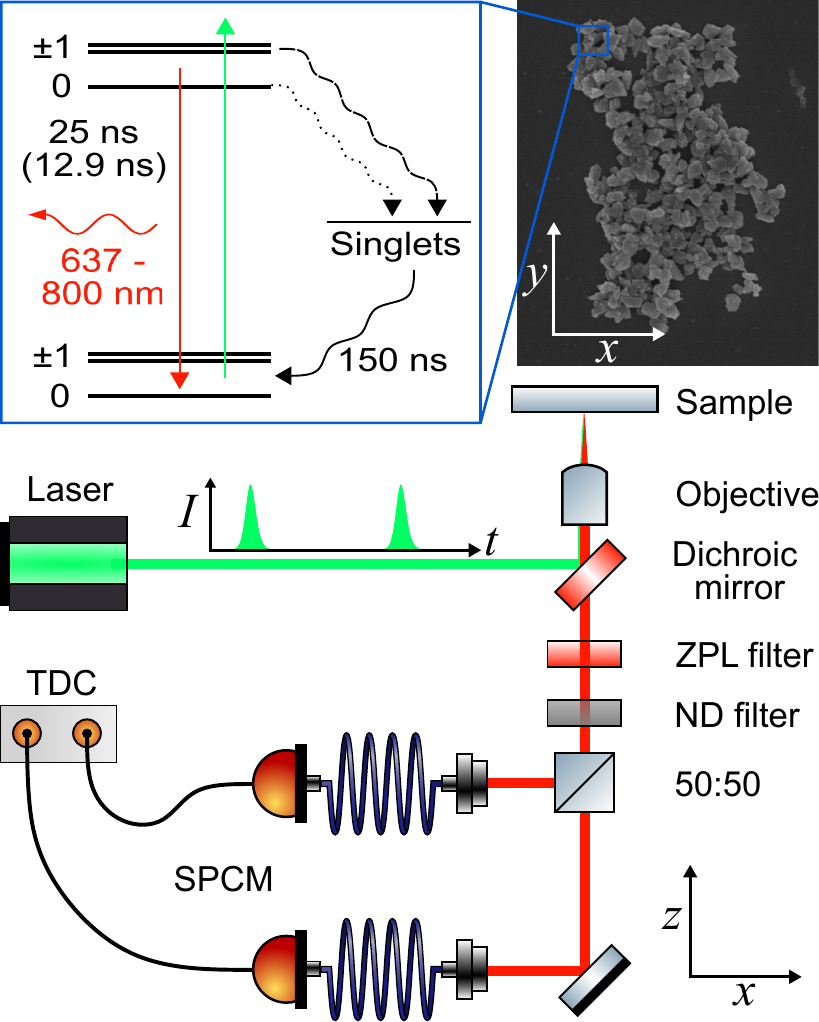}
\caption{Energy-level system of the NV center in diamond and experimental setup consisting of a confocal microscope and a HBT interferometer. Agglomerates exceeding a size of $A>\SI{1}{\micro\meter^2}$ are raster scanned in the $xy$-plane.}
\label{fig:setup}
\end{figure}

The negatively charged nitrogen-vacancy center is a paramagnetic point defect in the diamond lattice composed of a substitutional nitrogen atom (N) and an adjacent vacant lattice site (V) \cite{Doherty.2013}. 
This defect features local energy levels in the diamond's band-gap, which can be addressed with optical and microwave wavelengths, as depicted in figure \ref{fig:setup}. 
The electronic ground state 
and optically excited state 
are spin-triplet states $(m_s=0, \pm1)$ with separation energy of \SI{1.945}{e\volt} corresponding to a zero phonon line (ZPL) at $\lambda_{\mathrm{ZPL}}=\SI{637}{\nano\meter}$, followed by phononic bands of vibrational modes. The lifetime of the 
optically excited state differs for NV centers in nanodiamonds ($\tau_{\mathrm{ND}} \approx  \SI{23}{\nano\second} - \SI{25}{\nano\second}$) and bulk diamond ($\tau_{\mathrm{bulk}} = \SI{12}{\nano\second} - \SI{13}{\nano\second}$) \cite{Collins.1983,Batalov.2008,Beveratos.2001,Neumann.2009}.
Further, a second decay path via intersystem-crossing (ISC) into long-lived ($\tau_{\mathrm{ sing}} \approx \SI{150}{\nano\second}$ \cite{Robledo.2011}) singlet states 
and a subsequent second ISC back to the triplet ground state exists. The $m_s=\pm1$-states preferrably decay via this path, and therefore, optical pumping leads to spin polarization of the $m_s=0$ ground state.

The experimental setup consists of a home-built confocal microscope, as depicted in figure \ref{fig:setup}. NV centers are optically excited by an off-resonant pulsed laser source featuring a wavelength of $\lambda_L=\SI{517}{\nano\meter}$, pulse-widths of \SI{500}{\pico\second}, repetition rates of \SI{5}{\mega\hertz}, and a focus beam diameter of approximately \SI{350}{\nano\meter} on the sample plane. 
The sample consists of highly doped nanodiamonds in an aqueous solution \cite{.17.07.2018}. They have an average size of \SI{100}{\nano\meter} and more than \SI{1000}{} NV centers per single crystal. We drop cast them as received on a glass substrate and dry the sample with a contact heat plate to vaporize the water.
Fluorescence radiation of NV centers is collected by the same microscope objective transmitted through a dichroic mirror and wavelength-filtered by a bandpass to transmit a small band of \SI{637\pm 3}{\nano\meter}. 
After that, the fluorescence light is divided up at a 50:50 beam splitter and fiber-coupled to two single-photon counting modules (SPCM), forming a Hanbury-Brown and Twiss interferometer \cite{HANBURYBROWN.1956}. 
To reduce the total count rate to avoid measurements close to these detectors' saturation point, a neutral density (ND) filter ($\mathrm{OD}=2.0$) was added in front of the beam splitter.
A time-to-digital converter (TDC) tracks the arrival of single photons and calculates the average count rate as well as coincidences in each measurement. Appendix \ref{app:corr}
gives more details on the coincidence measurement and extraction of the pulsed second-order photon correlations.

To observe lifetime variations and photon bunching associated with collective effects, we investigated 104 agglomerates of nanodiamonds with an expansion of $A<\SI{1}{\micro\meter^2}$ and 12 agglomerates exceeding this size.
For agglomerates within $A<\SI{1}{\micro\meter^2}$  being in the order of the laser focus, we raster-scan the sample in $x$- and $y$-direction with a step-size of $s=\SI{0.3}{\micro\meter}$, choose the position of maximum count rate, and perform a single measurement. 
We group these structures into three different size categories according to the size evaluated from SEM images. These three categories are $A<\SI{0.05}{\micro\meter^2}$, $\SI{0.05}{\micro\meter^2}<A<\SI{0.2}{\micro\meter^2}$, and $\SI{0.2}{\micro\meter^2}<A<\SI{1}{\micro\meter^2}$.
For agglomerates exceeding a size of $A>\SI{1}{\micro\meter^2}$, we measure at every position of a raster scan with a step size of $s=\SI{0.4}{\micro\meter}$. 
We further increase the distance between sampling points to $s=\SI{1}{\micro\meter}$ for agglomerates exceeding $A>\SI{10}{\micro\meter^2}$ to reduce the total measurement time. 
This procedure allows quantifying variations of the lifetime and collective emission throughout agglomerates. 
Each measurement sequence consists of $1\cdot10^9$ repetitions of the laser pulse. During the first $1\cdot10^4$ repetitions, photon detection is shut to reach a steady-state and spin-polarization into the $m_s=0$ state.

\section{Transition to bulk material}

\begin{figure*}
\includegraphics[width=0.9\textwidth]{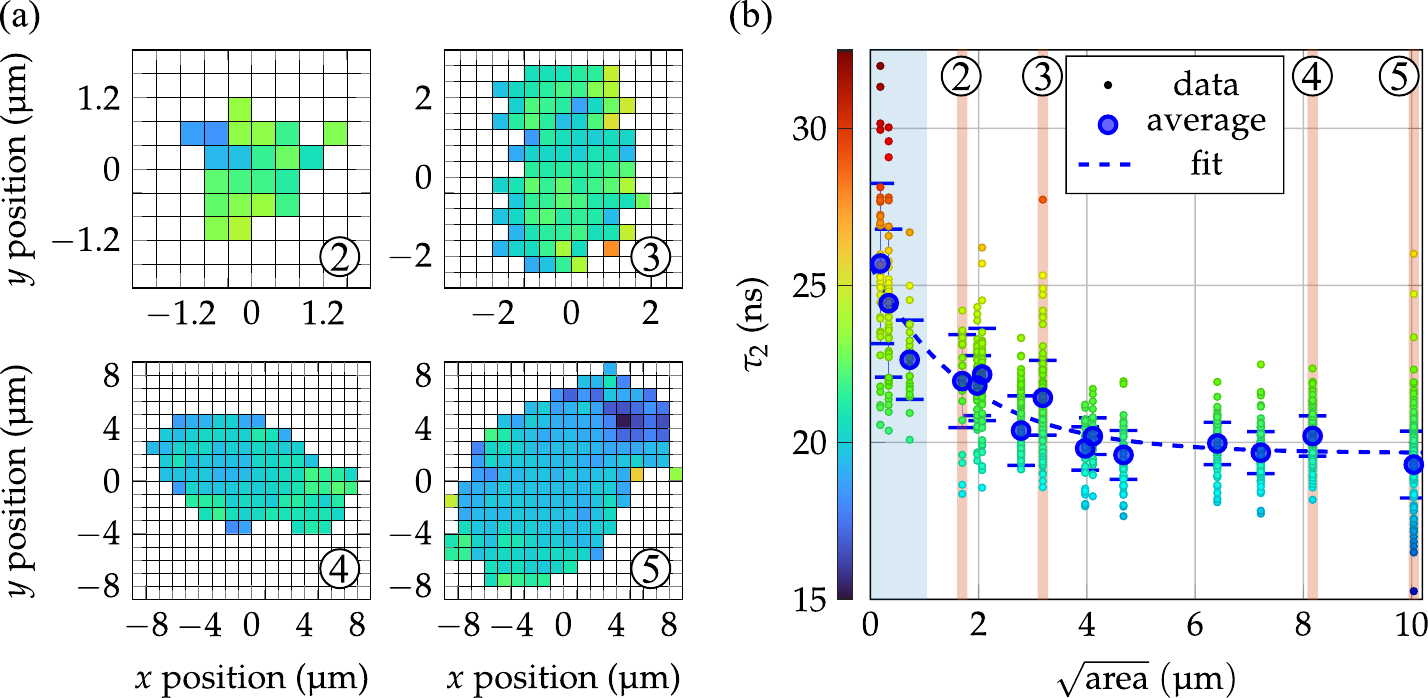}
\caption{Lifetime distribution as a function of agglomerate size. \textbf{(a)} Position-dependent fit value of slow decay $\tau_2$ in four different agglomerates with a common color bar. \textbf{(b)} Fit coefficients $\tau_2$ over the size-scale $l=\sqrt{\mathrm{area}}$ of the agglomerate they were measured in. The individual measurements are shown in the color bar used in (a). The region of single measurements without raster scan is shaded in blue, and the four agglomerates shown in (a) are highlighted and numbered according to figure \ref{fig:intro}. The data for each measurement is shown in appendix \ref{app:sND}. 
The count rate weighted average (standard deviation) is shown in thicker blue dots (bars) and was fitted by a phenomenological exponential decay.
}
\label{fig:LT}
\end{figure*}
We trace the transition from optical nano- to bulk-like properties of NV centers by scanning agglomerates with different sizes and performing a fluorescence lifetime measurement (FLIM), as described in section \ref{ch:setup}.
To quantify changes between samples with different size as well as variations within agglomerates, the measurement data is approximated by a model considering the time-dependent excitation of the laser pulse and a subsequent bi-exponential decay according to
\begin{equation}
   I(t) = a_1 \exp{(-t/\tau_1)} + a_2 \exp{(-t/\tau_2),}
    \label{eqn:biexp_decay}
\end{equation}
where $I(t)$ is the time-resolved intensity detected, $a_1$ and $a_2$ are the amplitudes at time $t=0$ of a fast decay $\tau_1$ and slow decay $\tau_2$.
The latter value $\tau_2$ is associated with the transition to bulk-like optical emission due to a change of the density of states (DOS) given by the refractive index. 

In figure \ref{fig:LT}(a), we show the position-dependent slow decay component $\tau_2$ in four agglomerates with varying sizes. 
All measurement data of 104 small nanodiamond agglomerates within the three size categories and 12 raster scans of larger agglomerates are shown as a function of the agglomerate size-scale $l=\sqrt{\mathrm{area}}$ in figure \ref{fig:LT}(b). 
The area on the substrate's surface of each structure was extracted from SEM images. 
For the full raster scans, we consider the area of the respective agglomerate and evaluate those positions that show connected fluorescence above a threshold of \SI{36.8}{\percent} of the maximum count rate measured in the smallest agglomerate.

For nanodiamonds of the smallest size category, we find a lifetime ranging from \SI{21}{\nano\second} to \SI{33}{\nano\second} with a count rate weighted average of \SI{26}{\nano\second}, which is in good agreement to the previously reported optical lifetimes of \SI{23}{\nano\second} \cite{Neumann.2009} and \SI{25}{\nano\second} \cite{Beveratos.2001}.
With increasing agglomerate size, the variation within each data-set shows, in general, a decrease. The position-resolved scans  show that data points with a significantly higher lifetime than average are typically located at the edges of the agglomerates (cf. agglomerate 3).
The average of fitted $\tau_2$ coefficients shows a reduction from $\SI{26}{\nano\second}$ towards a value of $\SI{19.5}{\nano\second}$. 
We interpret this change as a transition to bulk-like optical properties due to the change of the local DOS via the effective refractive index.
We determine the length scale $l_{\mathrm{nb}}$, on which this transition occurs by an exponential fit $A\cdot \exp{-a/l_{\mathrm{nb}}}+c$ to the weighted average, with a decay constant $l_{\mathrm{nb}}=\SI{1.8(3)}{\micro\meter} = \SI{2.8(3)}{\cdot}\lambda_\mathrm{ZPL}$, where the error is given by the standard deviation of the fit coefficient and $\lambda_\mathrm{ZPL}=\SI{637}{\nano\meter}$ is the emission wavelength of the NV center's ZPL.

Even though the lifetime does not reach the NV center's bulk lifetime in the order of \SI{12}{\nano\second} to \SI{13}{\nano\second} which is widely reported in the literature \cite{Collins.1983, Batalov.2008}, we assume that a transition to bulk-like optical properties is observed. 
This assumption stems from the fact that the graphitic surface \cite{Reineck.2017} of nanodiamonds and small voids between neighboring nanodiamonds lead to a reduction of the effective refractive index in contrast to bulk diamond. 
Further, the SEM images taken after the measurements suggest that the expansion of agglomerates along $x$- and $y$-directions, i.e., the extension along the glass substrate's surface, is much bigger than the extension along the $z$-direction, i.e., the laser beam axis. 
The agglomerates shown in the data above indicate to consist of typically one or two nanodiamond layers. 
For the largest agglomerate, however, we observe additional crystals on top of these layers (see figure \ref{fig:intro}(a), right image), which was also evident by an increase in the count rate (see figure \ref{fig:intro}(b), right image).
In this agglomerate, we observe at the position of highest count rate a further significant reduction of $\tau_2$ to \SI{15.2}{\nano\second}. At the same time, the section without such an extension into $z$-direction shows lower count rates and $\tau_2=\SI{20}{\nano\second}$ both in the same order of the previously discussed large planar agglomerates.

\section{Collective effects \label{collective_effects}}

\begin{figure*}[]
\centering
\includegraphics[width=0.9\textwidth]{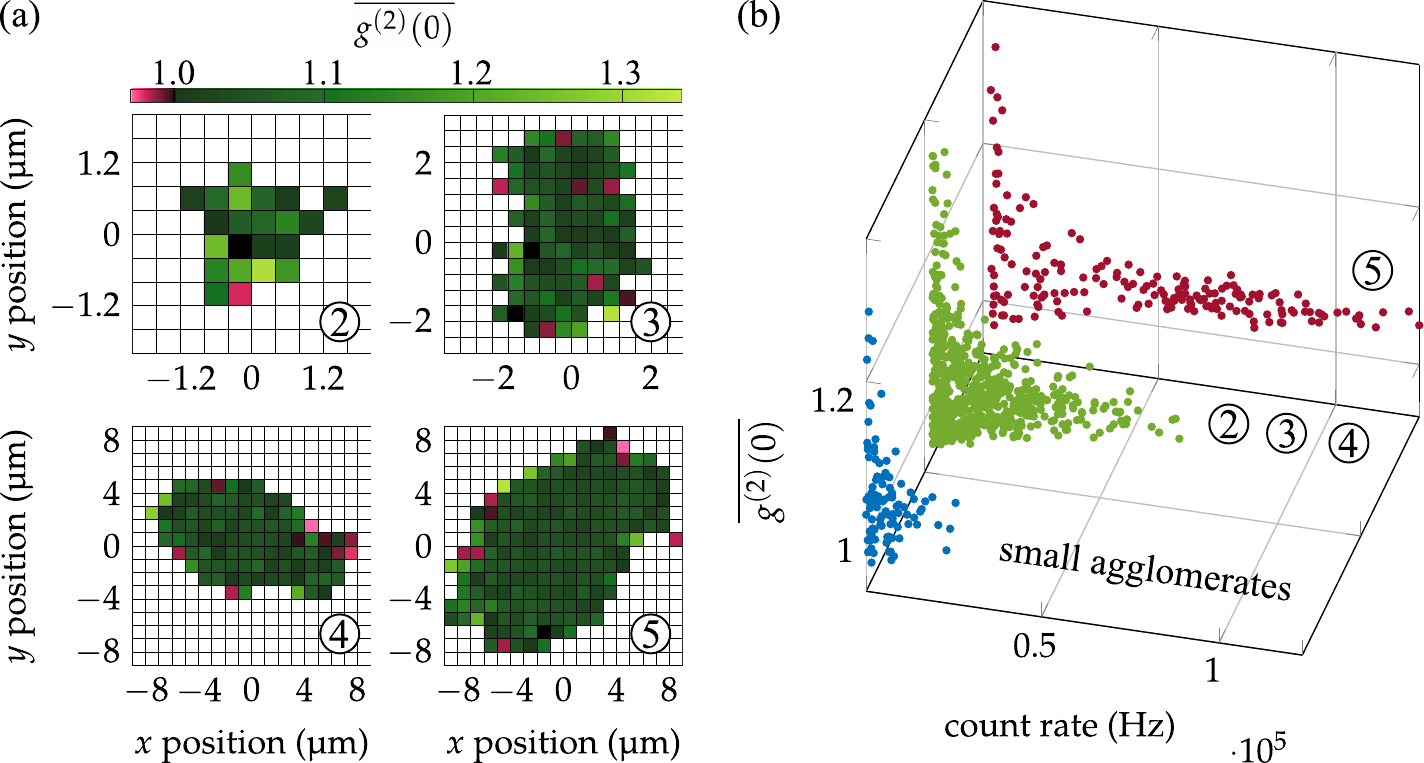}
\caption{\textbf{(a)} Position-dependent second-order correlation function with a common color bar. High photon bunching is observed in small agglomerates and on the edges of large agglomerates. Central sections of large agglomerates show almost constant photon bunching of $\overline{g^{(2)}(0)}=\SI{1.06}{}$. \textbf{(b)} Second-order photon correlation as a function of the count rate measured at the same time for agglomerates smaller than \SI{1}{\micro\meter^2} (blue), planar agglomerates (green) including agglomerates labeled as 2,3,4 in (a), and the largest agglomerate (red) labeled as 5.}
\label{fig:correlations}
\end{figure*}

Collective emission can be observed in different observables. First, a macroscopic spin emitting features a larger emission rate and hence a reduced lifetime \cite{Dicke.1954,Gross.1982}. Second, for collectively emitting NV centers, the statistics of photons detected is expected to change, which can be observed in the second-order correlation function as enhanced correlation \cite{Bradac.2017}. We find, however, that the lifetime measurements, which we show in detail in appendix \ref{app:fast_decay}, do not allow differentiating between the influence of dark decay \cite{Inam.2013, Orwa.2000, McCloskey.2014} and collective emission \cite{Bradac.2017}. We, therefore, investigate the second-order correlation function in detail.

\subsection{Second-order correlation function}

The second-order correlation at zero time delay $g^{(2)}(0)$ can be written via the photon distribution as \cite{Walls.2008}

\begin{equation}
    g^{(2)}(0)= \frac{\langle \hat{a}^+ \hat{a}^+ \hat{a} \hat{a} \rangle}{\langle \hat{a}^+ \hat{a} \rangle \langle \hat{a}^+ \hat{a} \rangle } =1+\frac{ \Delta n^2 -\langle n \rangle}{\langle n \rangle ^2}, \label{eq:g2_nsr}
\end{equation}
where $\langle n \rangle = \langle \hat{a}^+\hat{a} \rangle$ is the expectation value and $ \Delta n^2  = \langle n^2 \rangle - \langle n \rangle ^2 = \langle (\hat{a}^+) ^2 \hat{a} ^ 2 \rangle + \langle \hat{a}^+\hat{a} \rangle -  \langle \hat{a}^+\hat{a} \rangle ^2$ is the variance of the photon number expressed with creation and annihilation operators $\hat{a}^+$ and $\hat{a}$.
Equation (\ref{eq:g2_nsr}) shows that the second-order correlation depends on the first and second moment of the photon distribution. 
The measurement of the second-order photon correlation with a Hanbury-Brown-Twiss interferometer in the time domain is a standard tool to observe collective effects. 
We measure it after pulsed excitation and approximate the value by integration over a time window of $t'=\SI{0.5}{\nano\second}$ as explained in detail in appendix \ref{app:corr}. 
We refer to this value as $\overline{g^{(2)}(0)}$.
We expect to see photon bunching as a hallmark of collective emission in the measurement due to super-Poissonian photon statistics ($ \Delta n^2  > \langle n \rangle $) \cite{Walls.2008}. 

The position-dependent measurements of $\overline{g^{(2)}(0)}$ are shown in figure \ref{fig:correlations} (a). 
We measure photon bunching ($\overline{g^{(2)}(0)}>1$) as well as photon anti-bunching ($\overline{g^{(2)}(0)}<1$) in all agglomerates. 
From these measurements, we see that the amount of photon bunching reduces and shows fewer fluctuations with increasing size.
All data points are presented as a function of the count rate measured at the same spatial position and time in figure \ref{fig:correlations} (b). We differentiate between the measurements of small nanodiamond agglomerates measured at a single position, planar agglomerates, and the largest agglomerate, which shows the most substantial lifetime reduction.
We observe higher photon bunching for lower count rates and a higher variance in the data.
In large structures, we find the strongest photon bunching located at the edges of the agglomerate where the count rate is low and almost constant values of $\overline{g^{(2)}(0)} \approx 1.06$ in the central planar sections where the count rate is high.
Such a photon bunching can be explained by forming either thermal emission or collective emission of quantum emitters. 

\begin{figure*}
\includegraphics[width=.9\textwidth]{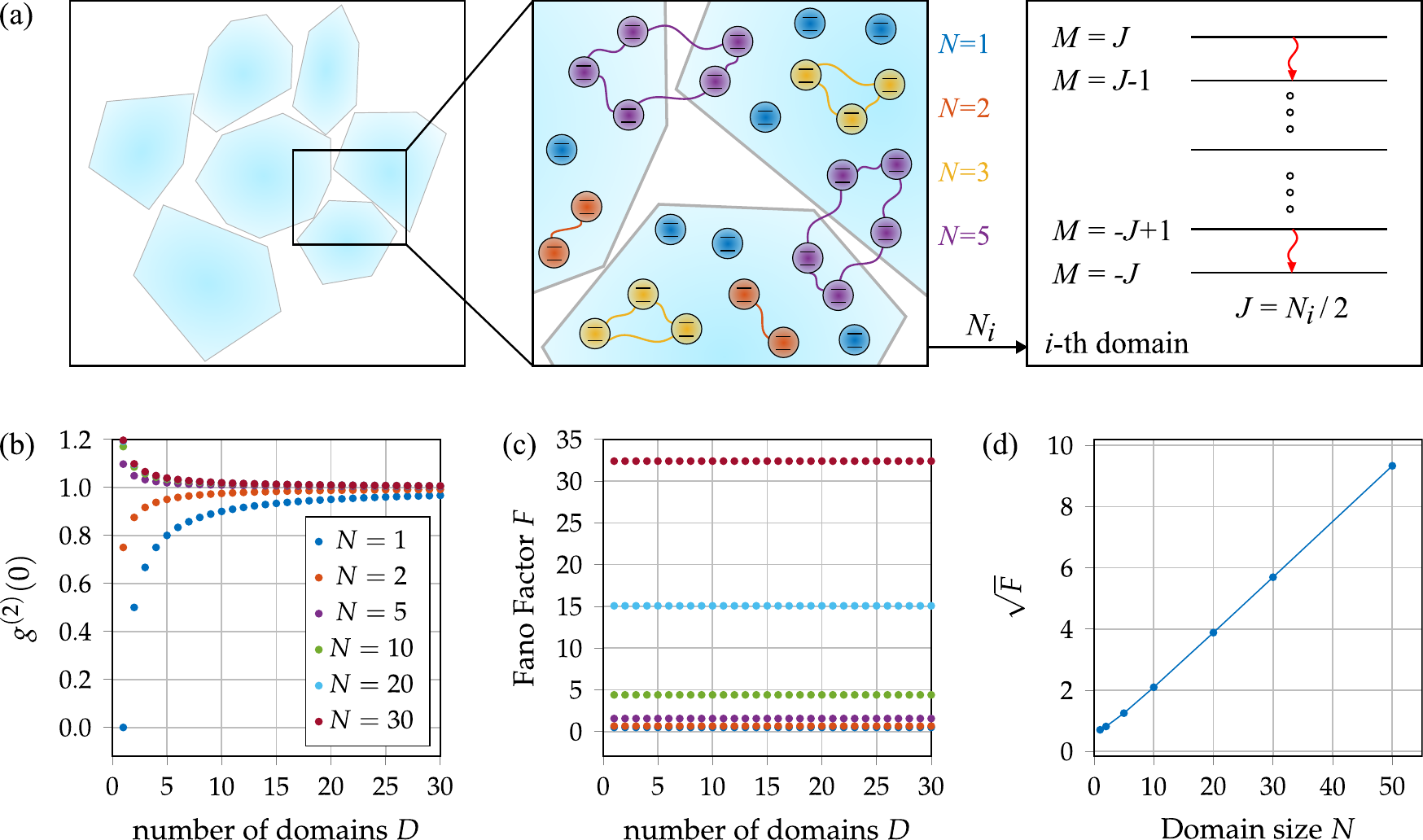}
\caption{\textbf{(a)} Model for collective emission in nanodiamond agglomerates. NV centers are assumed to form $i \in \{1,...,D\}$ spectral domains (shown in different colors) of $N$ two-level emitters. Each domain is described individually by Dicke-ladder-states.
\textbf{(b)} Calculated auto-correlation function $g^{(2)}(0)$ and \textbf{(c)} Fano factor $F$ as a functions of the total number of domains $D$ for different domain sizes $N$ under the assumption of a maximally mixed state. \textbf{(d)} Square root of Fano factor $\sqrt{F}$ as a function of the domain size $N$.}
\label{fig:ff_sim}
\end{figure*}

To differentiate between these different physical effects, we perform a second measurement of the second-order correlation in cw-mode for the largest agglomerate sample. 
A comparison of pulsed and cw second-order correlation is shown in appendix \ref{app:comp_pulsed_cw_corr}. 
In the cw-measurement, we observe no bunching or slight anti-bunching, expected from an ensemble of quantum emitters acting as individual emitters.
This discrepancy between pulsed and cw excitation is consistent with the interpretation of collective, superradiant effects. 
The formation of a collective spin is known to feature a finite coherence time. 
The pulsed measurement allows recording the correlation at very short times, smaller than the coherence time of a collective spin. By contrast, the cw excitation averages over long times, where a collective spin is expected to dephase and cease. 
In addition, the absence of photon bunching in cw-excited measurements also indicates the absence of chaotic or thermal emission. 
Moreover, we have found no reference reporting thermal emission of NV ensembles neither in nanodiamonds nor in bulk diamond at room temperature. 
Further, we have ruled out systematic errors by verifying the pulsed measurement scheme and evaluation on a nanodiamond with a low concentration of NV centers (see appendix \ref{app:control_sample}). 
We find similar results for pulsed and cw measurement of the second-order correlation function in this control sample. 
Therefore, we conclude that on short time scales, a collective emission of NV centers is taking place in nanodiamonds with high NV concentrations. 
We mention here that the observation of super-Poissonian photon statistics does not rule out the presence of an additional dark decay channel.

To obtain further insight, we adopt a model of collective emission in single-nanodiamond crystals \cite{Bradac.2017} and apply it to the agglomerates of nanodiamonds investigated here.
We assume the nanocrystals to consist of many spectral domains, where each domain acts collectively, but different domains emit independently from each other. A sketch of the model is presented in figure \ref{fig:ff_sim} (a).

In this model, an important figure of merit is the collective-domain size.
In a simple approach, one could interpret an increased photon bunching indicated by an increased second-order correlation as a sign of increased super-Poissonian emission due to larger collective domains. 
In this case, the observation of the highest photon bunching in individual nanodiamonds and at the borders of agglomerates would imply that collective domains are smaller in central positions of agglomerates.
However, this contradicts the observation of increased fast optical decay in larger agglomerates.

In our model, we assume multiple domains, where $D$ is the total number of domains contributing to the emission. 
For a single domain described by $N=2J$ Dicke-states $\ket{J,M}$, we calculate the second-order correlation and the expectation value $\langle n_N \rangle$ and variance $ \Delta n_N^2 $ of the photon number. 
The calculation and more detail on the model are presented in appendix \ref{app:sd}. 
We furthermore assume that single nanodiamonds and agglomerates comprise many domains, where each domain acts independently from all other domains, i.e., their photon distributions are statistically independent. Each domain $i=1,...,D$ is described by the respective expectation value $\langle n_{N_i} \rangle$ and variance $\Delta n_{N_i}^2 $ of the individual domain's photon number.
Thus, we calculate the total photon distribution's first and second moment of multiple domains as the sum of individual domains according to
\begin{equation*}
    \langle n \rangle = \sum_{i=1}^D \langle n_{N_i} \rangle
\end{equation*}{}
and
\begin{equation*}
    \Delta n^2  = \sum_{i=1}^D  \Delta n_{N_i}^2 .
\end{equation*}{}
The second-order correlation at zero time delay of $D$ domains with different domain sizes $N_i$ can be calculated using equation (\ref{eq:g2_nsr}). 
Assuming an initial maximally mixed state $P_{J,M}(0)=1/(N+1)$ \cite{Bradac.2017}, we have calculated the correlation function of different domain sizes $N$ over the total number of such domains $D$, as depicted in figure \ref{fig:ff_sim} (b).

The values calculated for $D$ domains of independent quantum emitters with a domain size $N=1$ yield the well-known relation of $g^{(2)}(0)=1-1/D$.
However, $g^{(2)}(0)$ depends on the domain size $N$ as well as the number of such domains $D$. Therefore, the second-order correlations approach $g^{(2)}(0) \approx 1$ for a large number of independent domains $D$ for all domain sizes $N$.

In the experiment with various agglomerate sizes, the number of domains excited by the laser beam has high variations. Therefore, a simple comparison of measured correlation functions between such different samples can not provide complete information on the average collective domain sizes. Consequently, we establish an alternative method based on the fluctuations of the photon distribution in the following.

\subsection{Fluctuations}

To have a figure of merit that is (i) independent of the number of contributing domains $D$ and has (ii) a linear scaling for collective domain sizes $N$, we introduce the square root of the Fano factor
\begin{equation*}
   \sqrt{F}=\sqrt{\frac{ \Delta n^2 }{\langle n \rangle}}
\end{equation*}
for quantification.
The Fano factor $F$ measures deviations from a Poissonian distribution, where a value larger (smaller) than unity indicates super- (sub-) Poissonian fluctuations.
Based on the discussion above, we calculate and plot the Fano factor assuming a maximally mixed state $P_{J,M}(0)=1/(N+1)$ in different domain sizes $N$ over the total number of such domains $D$ in figure \ref{fig:ff_sim} (c).
We find that the Fano factor is independent of the number of contributing domains $D$ and shows almost quadratic scaling with domain size $N$. 
We emphasize that the scaling is strongly dependent on the initial assumption on the Dicke-ladder state. For the occupation of the highest ladder state, linear scaling of the Fano factor with the domain size $N$ is calculated.
Considering the square root of the Fano factor $\sqrt{F}$, we find a quantity that shows linear scaling for the collective domain size $N$ while being independent of the number of contributing domains $D$ as depicted in figure \ref{fig:ff_sim} (d).

\begin{figure}[]
\centering
\includegraphics[width=0.45\textwidth]{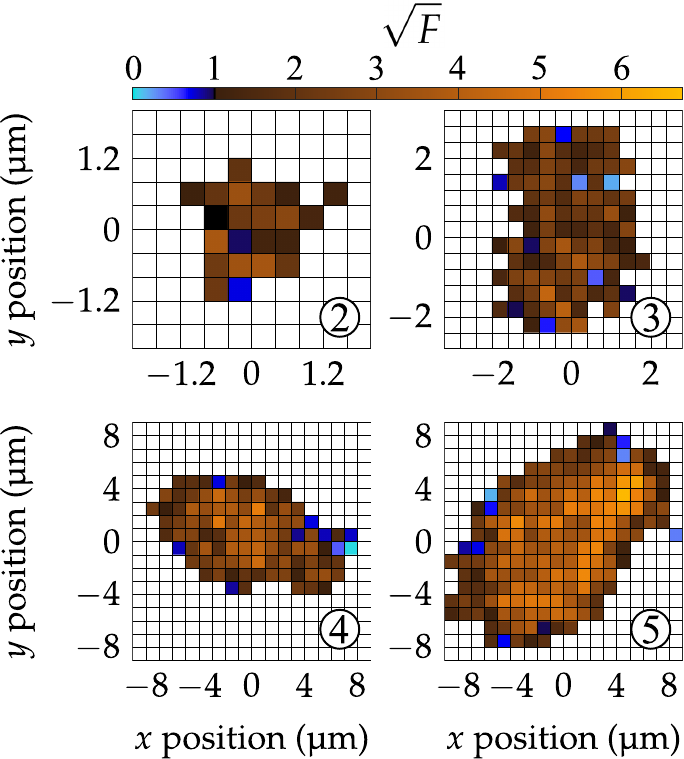}
\caption{Calculated sqare root of the Fano factor $\sqrt{F}$ as a function of position with a common color bar.}
\label{fig:fano_fac}
\end{figure}

However, the Fano factor is no direct measurement observable of the photon distribution. To access the Fano factor, we calculate it under the assumption that the expectation value of the photon number $\langle n \rangle$ scales linearly with the total count rate detected in a pulsed measurement.
We mention here that this approximation is only exact, assuming (i) a vanishing excitation pulse-width in time and (ii) $i=1,...,N$ independent quantum emitters each emitting with $\langle n_i \rangle = 1$. 
In the experimental sequence, the laser pulse has a width of \SI{500}{\pico\second}, which is significantly smaller than the typical lifetime of the NV center. 
This justifies the proportionality of the measured count rate to the number of NV centers excited.
However, for a collective domain, the mean photon number $\langle n \rangle$ will exceed the number of its constituents and scale quadratically during emission, as described by Dicke \cite{Dicke.1954}. 
Since small domain sizes were reported in \cite{Bradac.2017}, we expect the calculated square root of the Fano factor $\sqrt{F}$ to show a relatively small deviation. 
In this case, the calculated quantity of $\sqrt{F}$ can be understood as a quantitative measure for the formation of collective effects.

\begin{figure*}
\centering
\includegraphics[width=0.8\textwidth]{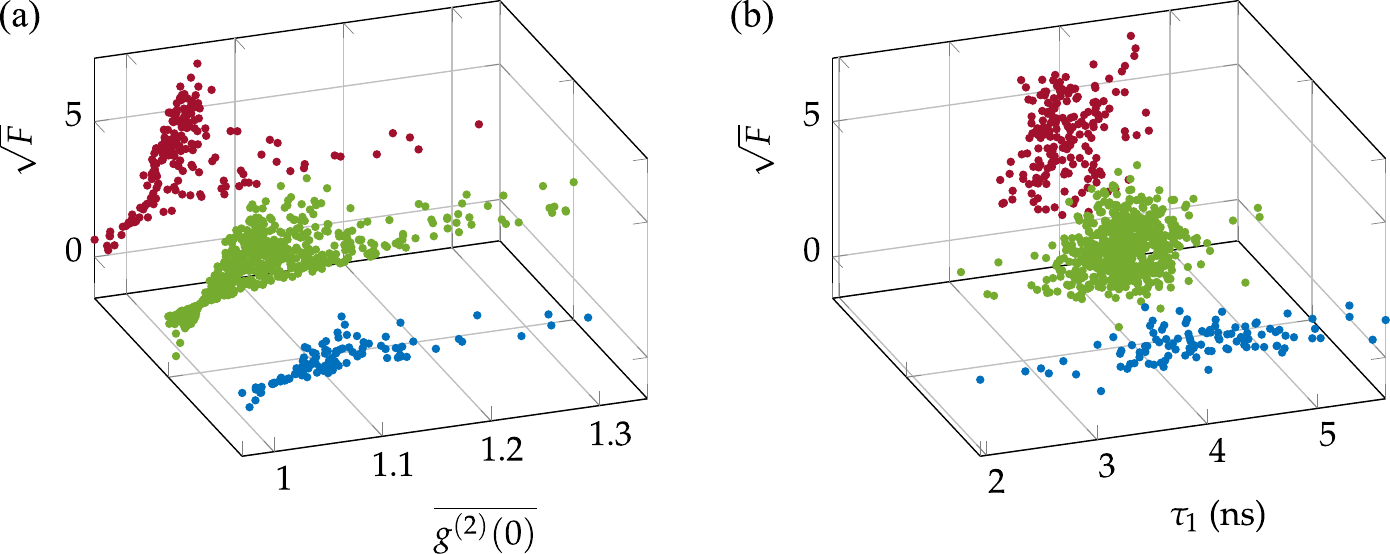}
\caption{Dependence of the square root of the Fano factor $\sqrt{F}$ on the \textbf{(a)} second-order correlation function $\overline{g^{(2)}(0)}$ and \textbf{(b)} fast-decay lifetime $\tau_1$ ; blue: small nanodiamonds with $A<\SI{1}{\micro\meter^2}$, green: planar agglomerates, red: largest agglomerate. Measurement errors are not shown to increase the visability. The statistical errors of the second order correlation function $\overline{g^{(2)}(0)}$ and square root of the Fano factor $\sqrt{F}$ are typically larger where the count rate is comparatively small, which is at high values of photon bunching.
}
\label{fig:ff_dep}
\end{figure*}

To estimate the mean photon number $\langle n \rangle$, we use the measured count rate and normalize it to the count rate measured with the same sequence for a nanodiamond of the control sample housing 10 NV centers shown in appendix \ref{app:control_sample}. 
This allows us to find a bias and scaling for the Fano factor and makes our results comparable with further studies in different experimental setups.
Moreover, we calculate the variance as $ \Delta n^2  = (g^{(2)}(0) - 1) \langle n \rangle ^2 + \langle n \rangle$ from the measured quantities of the count rate and second-order correlation function. 

Figure \ref{fig:fano_fac} shows the calculated square root of the Fano factor $\sqrt{F}$ as a function of the position for four agglomerates.
Our analysis leads to non-physical Fano factors $F<0$. 
However, the error given by evaluating the second-order correlation measurement allows $F>0$ for all those measurements.
In contrast to the measured second-order correlation function shown in figure \ref{fig:correlations} (a), we find an increase of the calculated square root of the Fano factor $\sqrt{F}$ with the agglomerate size. 
We observe small values of $\sqrt{F}$ in all agglomerates located on the edges and an increasing modulus towards central positions.
From these measurements, we deduce that the average collective domain size, which scales at least linearly with $\sqrt{F}$, increases in nanodiamond agglomerates. 
For the largest agglomerate, we find maximum $\sqrt{F}=\SI{6.75}{}$ at the position of maximum count rate, minimum $\tau_2$ lifetime coefficient, and extension into the $z$-direction. 
This value exceeds the maximum of nanodiamonds of the smallest size category by a factor of \SI{2.5}{}. 
The discussion above suggests that the average collective domain size increases at least by the same amount.
Further, large values of $\sqrt{F}$ indicating larger collective domains are observed in regions of high nanodiamond density. They might suggest a formation of such domains over multiple nanocrystals nearby.

In addition, the square root of the Fano factor $\sqrt{F}$ allows drawing wide-reaching conclusions on the emission properties  compared with other observables.
Therefore, we evaluate $\sqrt{F}$ for each spatial measurement position as a function of other measurement quantities, i.e., the fit coefficient $\tau_1$ of the optical decay and the second-order correlation function. 
Here, we differentiate between three classes: individual nanodiamonds, planar agglomerates, and the largest agglomerate with larger extension into the z-direction, as shown in figure \ref{fig:ff_dep}.

As described above, the Fano factor is calculated from the count rate and time-integrated second-order correlation function $\overline{g^{(2)}(0)}$. 
The highest values of the second-order correlation functions are measured in single nanodiamonds and small agglomerates. However, they do not coincide with the locations of the highest calculated Fano factors.
This depicts the discrepancy between the measured photon bunching and the average collective domain size of the discussion above. 
The maximum square root of the Fano factor $\sqrt{F}=6.75$ is found at $\overline{g^{(2)}(0)} = 1.07$ in the largest agglomerate. 
In each class, we observe a larger photon bunching in the presence of a smaller count rate leading to smaller values for $\sqrt{F}$ indicating smaller collective domains.

Further, prominent collective effects, i.e., high Fano factors, are observed in agglomerates where the fast decay component is in the order of $\tau_1 = \SI{4}{\nano\second}$ in all three classes.
These measurement results show that the measured fast decay component in the order of $\tau_1=\SI{4}{\nano\second}$ can be attributed to collective emission of NV centers. 
The increasing amplitude of this component in larger agglomerates shown initially in figure \ref{fig:intro} and discussed in detail in appendix \ref{app:fast_decay} 
is consistent with observing a larger average collective domain size.

\section{Conclusion}

In conclusion, we have traced the quantum-optical emission properties of NV ensembles in doped nanodiamonds of different agglomeration states. 
We observe the transition to continuous, bulk-like fluorescence-emission properties with increasing agglomerate size to occur on a length scale of $\SI{1.8}{\micro\meter}=2.8\cdot \lambda_{\mathrm{ZPL}}$. 
Furthermore, we observe collective, superradiant emission in pulsed-measurement sequences. While the second-order correlation function does not yield a clear signal of collective effects, introducing the Fano factor as a novel quantity based on the fluctuations of the photon statistics, we reveal superradiant emission even for an ensemble of collective domains with varying size.. 
We observe high Fano factors in the emission of NV centers in agglomerates, where a high amount of nanodiamonds is illuminated within the region of the laser focus, which might stem from the formation of collective domains over multiple nanocrystals.
Further, the collective emission was attributed to an additional fast optical decay in the order of $\tau_1 \approx \SI{4}{\nano\second}$.

The observation of collective emission in a poly-crystalline solid-state system at room temperature paves the way towards application of superradiance in a robust and versatile system. 
Further steps will include studies with varying excitation pulse widths and amplitudes to shed light on initial Dicke-state formation and the dephasing of the macroscopic collective spin. 
Spatial information about the size of collective domains can be accessed by a change of the excitation region either via super-resolution techniques such as stimulated emission depletion (STED) spectroscopy \cite{Rittweger.2009,ArroyoCamejo.2013} or via changes in the optical setup towards larger/smaller beam waists. 
Furthermore, deterministic in-situ positioning of individual nanodiamonds in arrays of optical tweezers \cite{Horowitz.2012, Neukirch.2013, Juan.2017} will allow to tailor the mutual distance, and thereby to compare inter-crystal collective effects with the properties of individual nanodiamonds. 
The integration of NV centers into a photonic environment featuring the coupling to a single optical mode such as optical cavities \cite{Wolters.2010} and waveguides \cite{Burek.2012, Momenzadeh.2015, Mouradian.2015, Shi.2016, Landowski.2020} could allow observation and control of collective interaction on larger length-scales.
Combining such devices with other established methods such as spin-to-charge conversion of NV centers \cite{Shields.2015} could enable further control of the collective emission direction and coupling in a collective system on the meso-scale. 
Beyond the fundamental understanding, the highly entangled Dicke states involved render collective emission attractive for applications in quantum metrology \cite{Zhang.2014, Wang.2014, Paulisch.2019, Garbe.2020}.

\begin{acknowledgments}
JG acknowledges support from the Max-Planck Graduate Center.
This work was funded by the Deutsche Forschungsgemeinschaft (DFG) under project no. 454931666. 
\end{acknowledgments}

\bibliography{bibliography}

\appendix

\begin{widetext}
\section{Correlation function measurements} \label{app:corr}
\FloatBarrier

In order to approximate the second order correlation at zero time delay $g^{(2)}(0)$ in a pulsed measurement, we follow the method detailed in \cite{Bradac.2017}. 
In brief, we measure the time difference of coincidences $c(t)$ in an interval of $[\SI{-1.5625}{\micro\second},\SI{1.5625}{\micro\second}]$ showing in total \SI{15}{} repetitions of the pulse sequence with a bin-size of \SI{25}{\pico\second}. 
We fit all peaks separately with the function $f_i(t)=A \exp{(-|(t-t_i)/\tau|)}+c$ in order to find the center $t_i$ of the $i$-th peak.
Afterwards, a time window $t'$ centered on this respective local maximum is used to compare the coincidences per time window of the $0$-th peak with the other 14 peaks labeled as $[-7, ..., -1, 1, ..., 7]$. 
The second-order correlation function is then the integration over an infinitesimal time
\begin{equation*}
    g^{(2)}(0)=\lim_{t' \rightarrow 0} \int_{-t'/2}^{t'/2} \frac{c_0(t)}{1/n \sum_{i\neq0} c_i(t) } dt.
\end{equation*}
We approximate $g^{(2)}(0)$ using a finite time window $t'$ as 
\begin{equation*}
    g^{(2)}(0)|_{t'}= \int_{-t'/2}^{t'/2} \frac{c_0(t)}{1/n \sum_{i\neq0} c_i(t) } dt'.
\end{equation*}
and calculate the error via error propagation of the standard deviation of $c_i(t)$. 
Measurement data and the extracted correlation function over different widths of the time window $t$ are depicted in figure \ref{fig:coincidences_g2_slice} .

\begin{figure*}
\centering
\includegraphics[width=.9\textwidth]{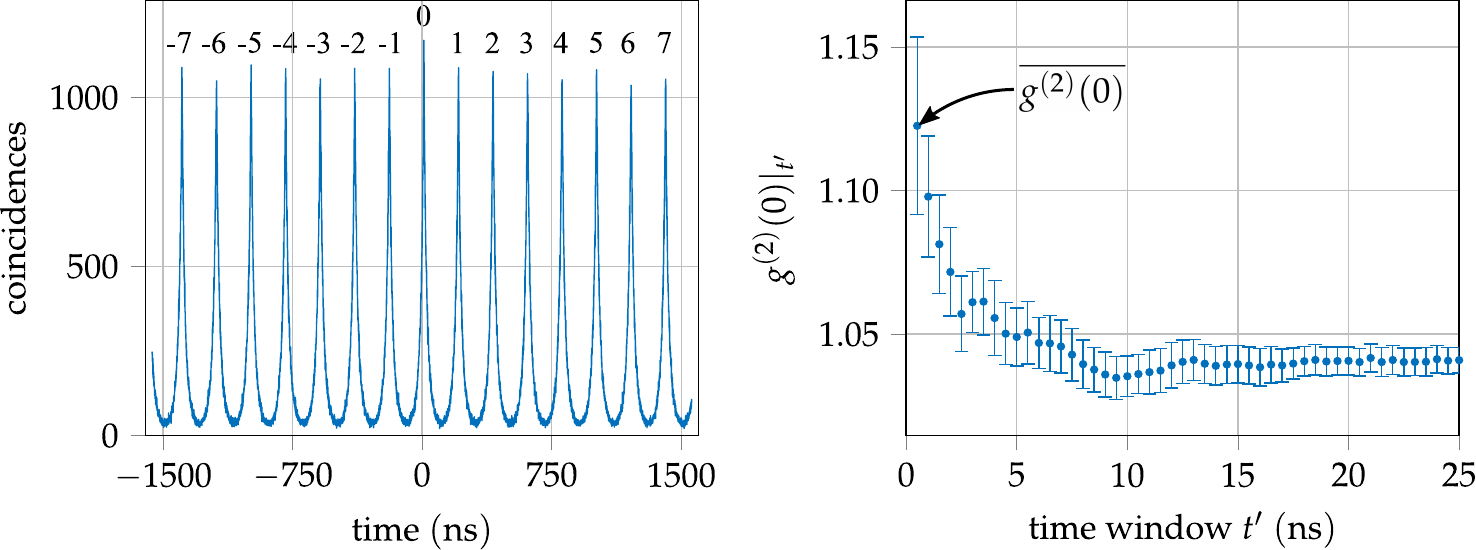}
\caption{\textbf{(a)} Measured coincidences over a time interval of 15 sequence repetitions. The "zeroth" peak shows the highest coincidences, i.e. photon bunching. \textbf{(b)} Approximated second order correlation using time slices and the normalization method described in the text.}
\label{fig:coincidences_g2_slice}
\end{figure*}

Ever decreasing time windows lead to a more precise approximation of the limit value but vice versa to an increased uncertainty, stemming from the higher variance of coincidences for smaller time windows of peaks $[-7, ..., -1, 1, ..., 7]$. The time window chosen to approximate the value of $g^{(2)}(0)$ is $t'=\SI{0.5}{\nano\second}$ in accordance to reference \cite{Bradac.2017}. We refer to this approximated value as $\overline{ g^{(2)}(0)}=g^{(2)}(0)|_{\SI{0.5}{\nano\second}}$ in the main text.

\section{Small nanodiamond agglomerates} \label{app:sND}

The individual measurements of nanodiamonds where a single masurement at maximum count rate was taken instead of a full confocal scan are presented in figure \ref{fig:sND}.

\begin{figure*}[h!]
\centering
\includegraphics[width=.9\textwidth]{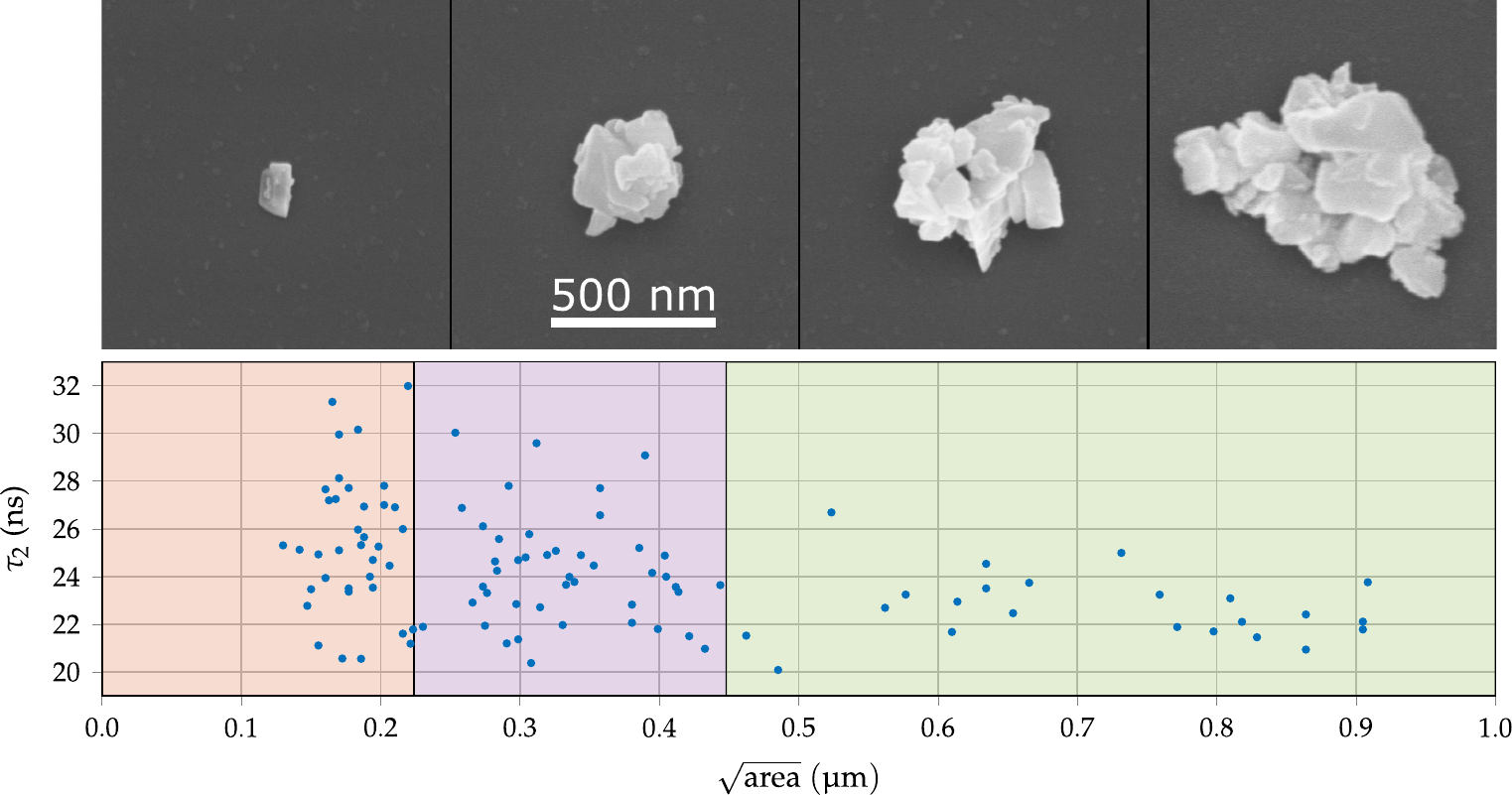}
\caption{\textbf{(a)} SEM images of different nanodiamond agglomerates with varying size. All images have a common scale. \textbf{(b)} $\tau_2$ lifetime from a bi-exponential fit of the lifetime measurtement. The size categories chosen are highlighted in different colors. The SEM images in (a) are typical candidates for the three categories, where the last two are both located in the biggest category.}
\label{fig:sND}
\end{figure*}

\section{Fast optical decay} \label{app:fast_decay}

In the literature the additional fast optical decay is attributed to several processes. These include collective effects \cite{Bradac.2017}, dark decay channels created by impurities and surface charge traps \cite{Inam.2013, Orwa.2000, McCloskey.2014}, as well as different ISC rates from the optical excited state's spin projections $m_s=0$ or $m_s=\pm1$ \cite{Neumann.2009, Batalov.2008, Robledo.2011}. 
The extracted fit-coefficients of the fast decay lifetime $\tau_1$ as well as the amplitude $a_1$ of the fit model (equation \ref{eqn:biexp_decay} in the main text) for the four agglomerates discussed in the main text is shown in dependence of the position in figure \ref{fig:LT2} (a) and (c). All data are shown in dependence of the agglomerate size in figure \ref{fig:LT2} (b) and (d).

\begin{figure*}[h!]
\includegraphics[width=0.9\textwidth]{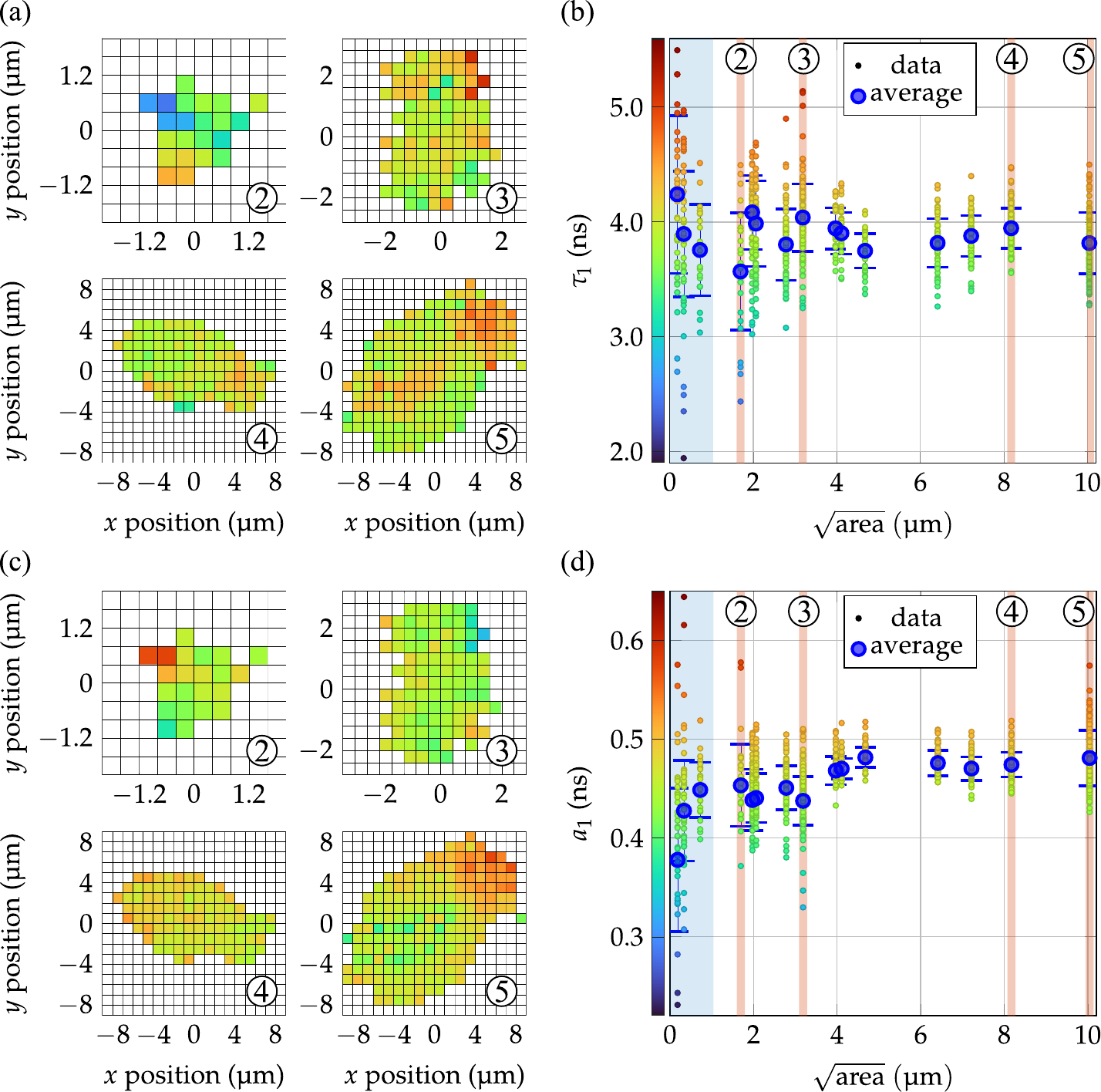}
\caption{\textbf{(a)} Fast decay lifetime $\tau_1$ of the fit model in dependence of the spatial position for the four agglomerates discussed in the main text and \textbf{(b)}  coefficients $\tau_1$ of all agglomerates in dependence on the agglomerate size measured in. \textbf{(c)} Fast decay fit amplitude $a1$ in dependence of the spatial position for the four agglomerates discussed in the main text and \textbf{(d)} in dependence of the agglomerate size.} 
\label{fig:LT2}
\end{figure*}

We observe on average $\tau_1 \approx \SI{4}{\nano\second}$ in all agglomeration-states. The variation in the data is decreasing for larger agglomerates. 
In the largest agglomerate, we observe higher values of $\tau_1$ at positions of extension into the $z$-direction.
Such small lifetimes are far below the reported decay of $m_s=\pm1$-states in nanodiamonds (\SI{12.8}{\nano\second} \cite{Neumann.2009}) as well as bulk diamond (\SI{7.3}{\nano\second} - \SI{7.8}{\nano\second} \cite{Batalov.2008, Robledo.2011}). 
Therefore, we rule out that the fast decay $\tau_1$ stems from increased ISC of $m_s=\pm1$-states.
Further, an increase of $a_1$ on small length-scales within the small size categories of single scans is observed. 
This effect is also evident in figure \ref{fig:intro} (c) of the main text, where the zoom-in section compares the contribution of the fast decay and the smallest nanodiamond-agglomerates show less pronounced fast decay. 
Again, the variation in the data dereases with increasing agglomerate size. The largest agglomerate is an exception and shows higher amplitudes $a_1$ at those positions where we see an extension into the $z$ direction.

In general, the effect leading to the fast decay in the order of $\tau_1 \approx \SI{4}{\nano\second}$ is more prominent in agglomerates.
However, from the lifetime measurements we can not differentiate between the influence of two possible contributions, i.e. dark decay and collective emission. Therefore, we focus on the analysis of the second order correlation in the main text.

\section{Comparison of cw and pulsed second-order correlation} \label{app:comp_pulsed_cw_corr}
For the largest agglomerate discussed in the manuscript the correlation function in dependence of the position has been detected in cw-mode as shown in figure \ref{fig:g2c}(a) in a second measurement. The difference of results in pulsed and cw-mode is shown in figure \ref{fig:g2c}(b). 

\begin{figure*}[h!]
\includegraphics[width=.9\textwidth]{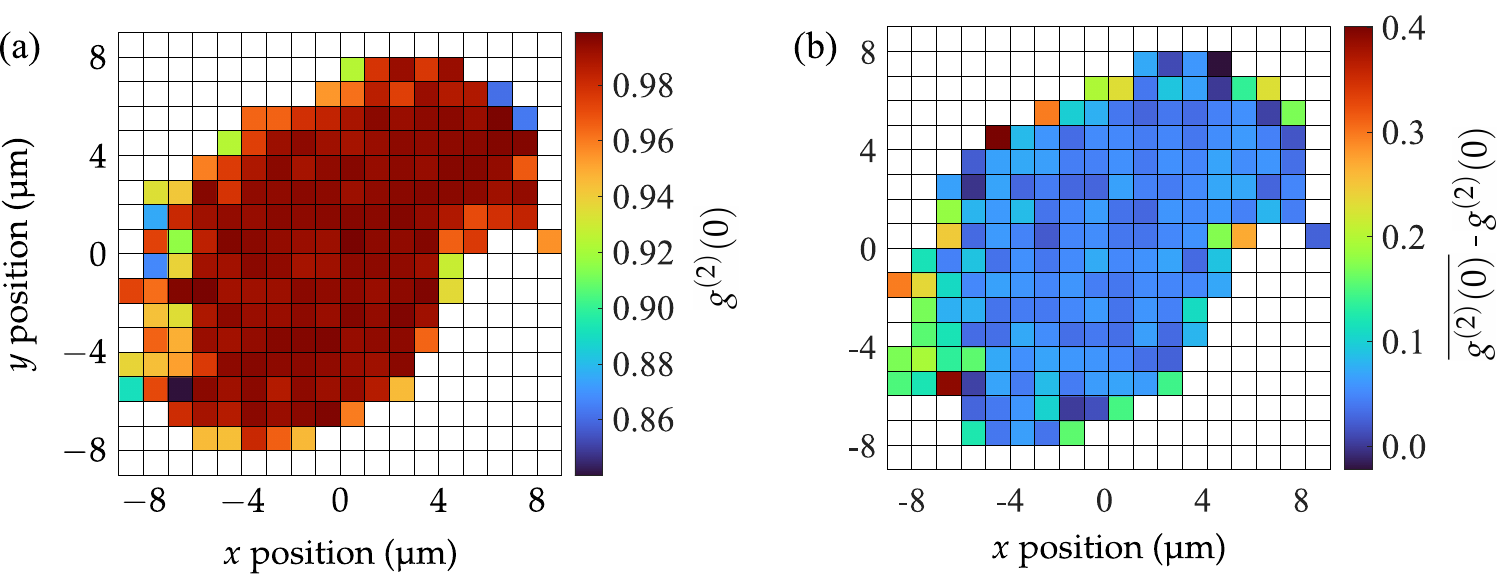}
\caption{\textbf{(a)} Measured cw second order correlation function at zero time delay $g^{(2)}(0)$ and \textbf{(b)} Difference of approximated second order correlation from the pulsed measurement (see figure \ref{fig:correlations} (a) in the main text) and cw second order correlation function $\overline{g^{(2)}(0)}-g^{(2)}(0)$ in dependence of the position.}
\label{fig:g2c}
\end{figure*}

As mentioned in the main text, we measure in general a cw second order correlation $g^{(2)}(0)<1$. The difference between cw and pulsed measurement shows a trend to larger values at positions of comparatively small count rate at the edges of agglomerates. 

\section{Control sample} \label{app:control_sample}

In order to rule out systematic errors of measurements performed with the pulsed sequence, nanodiamonds with few NV-centers but equal size were investigated with both measurement schemes. 
For these nanodiamonds we expect no formation of collective domains leading to anti-bunching and the agreement of both methods to infer the normalized second order correlation function $g^{(2)}(0)$. The integration time for both measurements has been increased but the applied pulse sequence has been left unchanged to all other pulsed measurements. 
The comparison of both measurements is depicted in figure \ref{fig:g2_10em}. The minimum value of the measured second order correlation function is shifted by approximately \SI{10}{ns} stemming from the different arm lengths of the HBT interferometer which employs differnt lengths of optical fibers.
For the investigated nanodiamond we find good agreement within the error bars of both measurement methods and conclude that this nanodiamond houses $10$ NV centers. 
From the measured count rate of those 10 NV centers, we infer the number of NV centers contributing in the agglomerate scans, in order to normalize the calculated Fano factor. 

\begin{figure*}[h!]
\includegraphics[width=.9\textwidth]{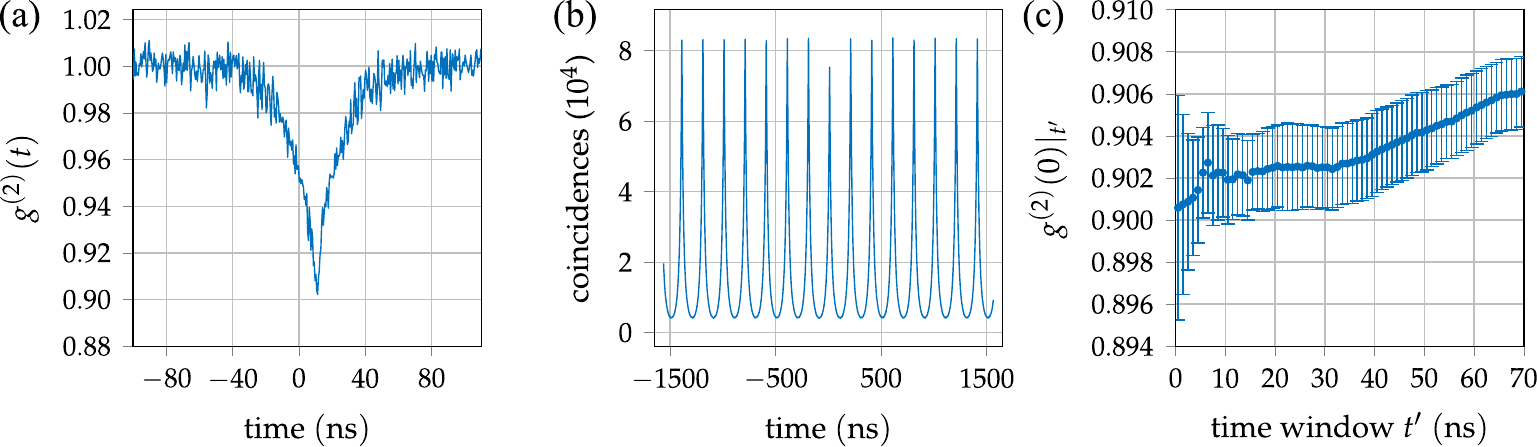} 
\caption{Comparison of measurement methods in nanodiamonds housing few NV centers. \textbf{(a)} Measured cw second order correlation function; \textbf{(b)} Coincidence measurement of pulsed excitation. \textbf{(c)} Approximated second order correlation at zero time delay for different time windows $t'$ of the pulsed measurement.}
\label{fig:g2_10em}
\end{figure*}

\section{Photon distribution of single collective domains}\label{app:sd}

In order to calculate the second order correlation we use the expectation value $\langle n \rangle$ and the variance $ \Delta n^2 $ of the photon number.
We model a single collective domain of $N$ emitters with Dicke-states $\ket{J,M}$ as a spin-ladder of spin $1/2$ systems, where $J$ describes the maximum total spin of a domain and $M \in  \{-J, -J+1, ..., J\}$ are the eigenvalues along the quantization axis, denoting the number of excitations in the system as depicted in figure \ref{fig:ff_sim} (a).

Since we are only interested in the second order correlation at zero time delay $g^{(2)}(0)$, we assume a vanishing excitation pulse to excite the emitters into the highest $J=N/2$-subspace and neglect decoherence processes of the collective spin to lower subspaces. 
In the model, we describe this by an initial population of these Dicke-states $P_{J,M}(0)$.
The collective state after excitation is then given by $\rho_N (0) = \sum_M P_{J,M}(0) \ket{J,M} \bra{J,M}$. 

For such a single Dicke-ladder of size $N$ we calculate the expectation value $\langle n_N \rangle$ as
\begin{flalign*}
    \hspace{0.1\textwidth} 
    & \langle n_N \rangle = \langle J^+J^- \rangle = A &&
\end{flalign*}
and the variance $ \Delta n_N^2 $ as
\begin{flalign*}
     \hspace{0.1\textwidth} 
     & \Delta n_N^2 = \langle J^+J^-J^+J^- \rangle + \langle J^+J^- \rangle - \langle J^+J^- \rangle \langle J^+J^- \rangle = B+A-A^2, &&
\end{flalign*}
where we introduced the quantities $A=\langle J^+J^- \rangle$ and  $B=\langle J^+J^-J^+J^- \rangle$.
We calculate $A$ and $B$ by taking the trace

\begin{flalign*}
    \hspace{0.1\textwidth}
    A  & = \mathrm{Tr}(\rho_N(0) A) && \\
       & = \sum_{M=-J}^{M=J} P_{JM}(0) \bra{J,M} J^+J^- \ket{J,M} \\   
       & = \sum_{M=-J}^J P_{J,M}(0)[J(J+1)-M(M-1)] \mathrm{,}
\end{flalign*}

\begin{flalign*}
    \hspace{0.1\textwidth} 
    B & = \mathrm{Tr}(\rho_N(0) B) && \\
      & = \sum_{M=-J}^{M=J} P_{JM}(0) \bra{J,M} J^+J^-J^+J^- \ket{J,M} \\
      & = \sum_{M=-J}^J P_{J,M}(0)[J(J+1)-M(M-1)][J(J+1)-(M-1)(M-2)].
\end{flalign*}
Entering $A$ and $B$ leads to the expectation value
\begin{flalign*}
    \hspace{0.1\textwidth}
    & \langle n_N \rangle = \sum_{M=-J}^J P_{J,M}(0)[J(J+1)-M(M-1)] &&\\
\end{flalign*}
and the variance
\begin{flalign*}
    \hspace{0.1\textwidth}
    \Delta n^2  & = \sum_{M=-J}^J P_{J,M}(0)[J(J+1)-M(M-1)][J(J+1)-(M-1)(M-2)] &&\\
    & + \sum_{M=-J}^J P_{J,M}(0)[J(J+1)-M(M-1)] \\
    & - (\sum_{M=-J}^J P_{J,M}(0)[J(J+1)-M(M-1)])^2.
\end{flalign*}
\end{widetext}

\end{document}